# Does the metallic 1T phase WS$_2$ really exist?

Yung-Chang Lin[1]*, Hideaki Nakajima[2], Chung-Wei Tseng[3], Shisheng Li[4], Zheng Liu[5], Toshiya Okazaki[2], Po-Wen Chiu[3], Kazu Suenaga[1]*

[1]National Institute of Advanced Industrial Science and Technology (AIST), Tsukuba 305-8565, Japan

[2]CNT-Application Research Center, National Institute of Advanced Industrial Science and Technology (AIST), Tsukuba 305-8565, Japan

[3]Department of Electrical Engineering, National Tsing Hua University, Hsinchu 30013, Taiwan

[4]International Center for Young Scientists (ICYS), National Institute for Materials Science (NIMS), Tsukuba 305-0044, Japan

[5]Inorganic Functional Materials Research Institute, National Institute of Advanced Industrial Science and Technology (AIST), Nagoya 463-8560, Japan

## Abstract

The electronic and optical properties of transition metal dichalcogenides (TMDCs) in distinctive phases, such as 1H, 1T, and 1T' phases, are of fundamental importance for variety of applications. The 1H phase has been understood as a direct bandgap semiconductor. On the other hand, the electronic properties of the 1T and 1T' phases remain controversy in the theoretical and experimental perspectives. In this study, we explore the optical properties of monolayer WS$_2$ in 1H, 1T, and 1T' phases using Raman and photoluminescence, corroborated with the atomic structure identified by scanning transmission electron microscopy. Despite of earlier theoretical investigations which all predict the *metallic* 1T phase, we experimentally discovered that the 1T-phase WS$_2$ is a direct band gap semiconductor and optically indistinguishable from the 1H phase. In a sharp contrast, the 1T'-phase WS$_2$ shows a gapless nature in its bandstructure with the quenched exciton transition as expected. Our experimental findings may give an interesting twist on the existing literatures reporting the metallic nature of 1T phase because the 1T phase has a strong tendency to co-exist with the metallic 1T' phase.

**Corresponding authors:** yc-lin@aist.go.jp, suenaga-kazu@aist.go.jp

The group VI layered transition metal dichalcogenides (TMDCs), such as $MoS_2$ and $WS_2$, have caught great research attentions due to the semiconducting characteristic and visible luminescence behavior in their monolithic flakes. [1–3] These properties lead to feasible applications in electronics and optoelectronics. Besides, their unique defects, dopants, and edge structures possess excellent electrocatalytic activities, showing great potential in solving energy related issues. [4,5] What brings TMDCs more conspicuous than other 2D materials such as graphene, BN, and phosphorene is that TMDCs possess more than one electronic property within their monolayers. [6] This is known as polymorphism, one of the most intriguing features of group VI TMDCs. Trigonal prismatic coordination, nominated as 1H phase, is the stable atomic configuration of semiconducting $WS_2$ for their bulk crystals and single-layers, in which the S atoms at top and bottom planes are aligned in *c*-axis forming a hexagonal arrangement with the W atoms at middle plane (Fig. 1a). When one of the S planes is displaced transversally to the hollow center of the hexagonal primitive cell with an octahedral symmetry, the electronic property is expected to dramatically convert from semiconducting to metallic, known as 1T phase (Fig. 1b). [7] Theoretically, the ground state energy of 1T phase is 85 meV higher than that of 1H phase. As a result, the metastable 1T phase will turn into 1H phase spontaneously under ambient condition, and to date, has never been found in natural bulk crystals yet.

It is known that charge plays an important role in the phase transition of group VI TMDCs. Chemical exfoliation via lithium intercalation was reported to transform the phase of TMDCs from 1H into 1T or 1T' phases. [8–10] The 1T' phase is a distorted 1T phase which exhibits a 2 x 1 superstructure and deforms into a dimerized zig-zag chain-like structure (see Fig. 1c). The chemically exfoliated monolayers often contain the mixture of 1T and 1T' phases which show additional peaks in Raman spectroscopy [11,12] and red-shift in the core-level binding energy [13,14] as compared to the 1H phase due to the existence of distinct symmetry. Although the metallic 1T phase had been discovered since the 1990s [15] and numerous potential applications, such as supercapacitors, hydrogen evolution catalysts, and memories, have been demonstrated to date, [16–19] the fundamental properties of 1T and 1T' are still vague and lead to intense arguments. For example: (1) In various of previous reports, the existence of 1T or 1T' phase was used to relied on the extra Raman peaks at lower energy (i.e., $J_1$, $J_2$, $J_3$ peaks) and the red-shift of binding energy. [11,12,19–22] However, the J peaks cannot originate from the 1T phase, which is actually from the 2 x 1 superstructure of 1T' phase who folds phonons at the edges of Brillouin zone into Γ and giving rise to new Raman modes. [12,23,24] This is also confirmed by recent growth of high crystallinity 1T' phase samples. [25,26] (2) Most of the literatures reported that both the 1T and 1T' phases are metallic behavior, [13,26] but several research groups held different opinions and reported 1T' as a semiconductor. [24,27,28] (3) The 1T phase has long been claimed to be very unstable and tends to transform into 1H phase at room temperature. In contrast, the 1H to 1T phase transformation has been directly visualized at high temperature (T=573K~973K) in a scanning transmission electron microscopy (STEM). [29]

Moreover, the 1T phase WS$_2$ can be directly synthesized on sapphire at 1123K by using chemical vapor deposition. [30] These discrepancies between the experimental results and the theoretical calculations point out that the properties of 1T and 1T' phases are still unclear and need more careful studies.

Hitherto, to confirm the atomic structures of TMDCs being 1H, 1T, or 1T' phases still rely on the direct atomic level visualization by STEM. The main obstacle to correctly assign the electronic and optical properties of TMDC monolayers to the corresponding polymorphic phases is due to the difficulty in collecting macroscopic measurements and local atomic structure information from the same area simultaneously. Here, we report a systematic study on the direct correlation of optical properties and atomic structures of single-layer WS$_2$ in 1H, 1T, and 1T' phases by combining Raman, photoluminescence (PL), exciton absorption, and STEM analyses. Strikingly, we found that the optical property of the 1T phase is nearly identical to the semiconducting 1H phase. Instead, the 1T' phase behaves quenching in the excitonic transition.

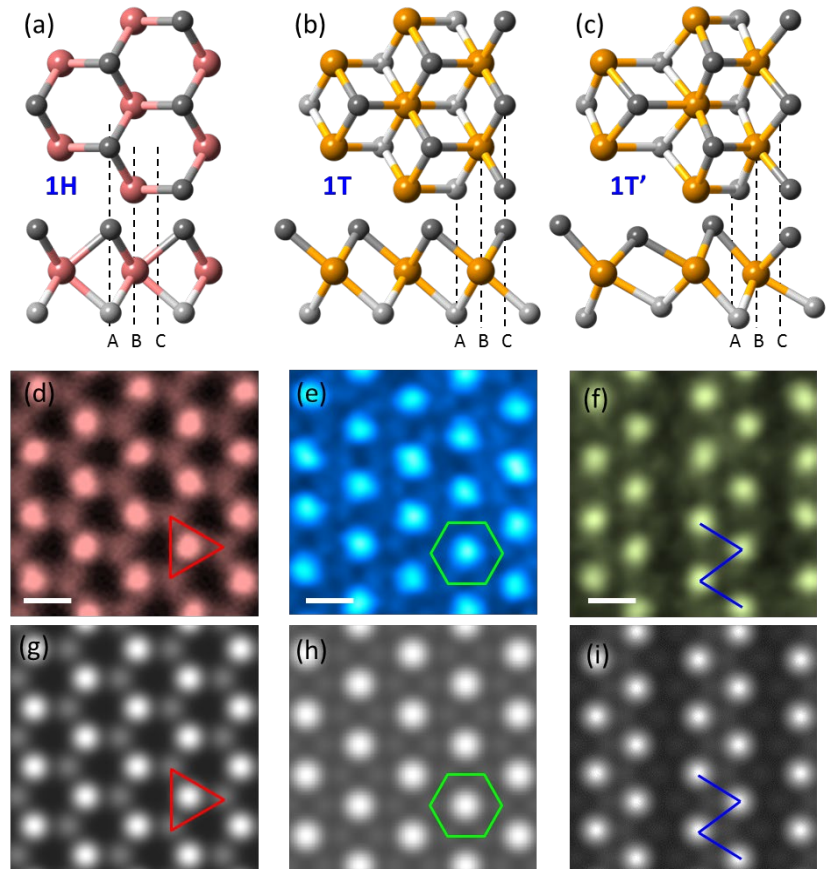

FIG. 1. (a-c) Atomic models of 1H, 1T, and 1T' phases. (d-f) Experimental ADF image of 1H, 1T, and 1T' phases. (g-i) Simulated ADF image of 1H, 1T and 1T' phases. Scale bars are 0.25 nm.

Figure 1d and 1g show the typical annular dark field (ADF) image and the simulation of 1H phase $WS_2$, where the overlapped S atomic planes near a W atom is arranged in triangular symmetry as highlighted by a red triangle. Figure 1e and 1h represent the experimental and simulated ADF images of 1T phase $WS_2$. The 6 nearest S atoms from a W atom are homogeneously arranged in a hexagonal shape highlighted by a green hexagon. Experimentally, imaging 1T phase by STEM requires precise adjustment of the microscope alignment. The ADF image of 1T phase can be strongly affected by aberrations such as coma, two-fold, and three-fold astigmatism. The three-fold astigmatism will result in one of contrast variation of the S atomic planes in the 1T phase; [31] on the other hand, the coma and the two-fold astigmatism result in distortion of the S atomic position in the 1T-phase. [30] Figure 1f and 1i show the ADF image and simulation of 1T' phase $WS_2$ where the dimerized W atoms deform into a zigzag chain-like structure and is highlighted by the blue lines.

In the previous work, butterfly-shape of monolayer $WS_2$ grown by chemical vapor deposition were reported consisting of the 1H phase in one wing and the 1T phase in the other. [30] Figure 2a shows an optical image of a $WS_2$ butterfly grown on sapphire and transferred to a $SiO_2$/Si substrate. The boundary between the two wings (assigned as A and B) is highlighted by a red dashed line. We performed the Raman and PL mappings on the $WS_2$ butterfly and show in Fig. 2b and 2c, respectively. The Raman mapping contour is plotted according to the $E_{2g}$ peak. The excitation was using green light laser at 532 nm with a power of 0.5 mW. One can see that, both the A and B domains show identical intensity contrast of the $E_{2g}$ mode. The extract Raman spectra of the A and B domains as well as additional characterization of lower energy and lower power excitation are shown in Fig. S1a and S1b. The PL mapping contour shown in Fig. 2c is plotted according to the integrated PL intensity within the energy range of 1.95 eV to 2.05 eV, which show also homogeneous contrast between the A and B domains. PL spectra extracted from various regions of the $WS_2$ butterfly showing very small variations are presented in Fig. S1c and S1d. There are several possible origins such as defects and impurities resulting in the PL intensity variation, [32] but still the left and right wings of the $WS_2$ butterfly behave like direct band gap semiconductor. We can firstly conclude that the phase of the WS$_2$ butterfly in the A and B domains is indistinguishable by the optical approaches. Therefore, the both domains could be simply assigned to 1H phase due to the semiconducting excitonic behavior. However, according to our careful repeated experiment on more than 50 WS$_2$ butterflies by using Raman, PL and precisely aligned STEM imaging (Fig. 2d, we found that one wing (A domain) is consisting of 1T phase (Fig. 2e), while the other wing (B domain) is consisting of 1H phase (Fig. 2f). We have ruled out the possibility of specimen tilt at the domain boundary (Fig. S2 and S3) or phase transformation during the TEM specimen preparation [30] and get another contrary inference that both 1T and 1H WS$_2$ are direct band gap semiconductor. In other words, the 1H and 1T phases WS$_2$ are indistinguishable by optical approaches, but the atomic-resolution STEM is the only way to differentiate. This rises the fundamental questions: which phase of WS$_2$ behaves metallic, and why 1T-phase WS$_2$ behaves direct band gap semiconductor.

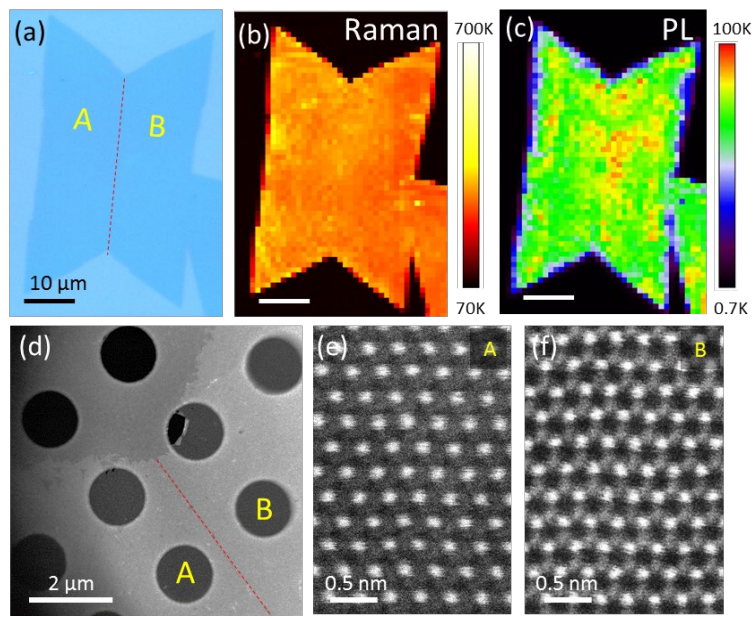

FIG. 2. (a) Optical image of a $WS_2$ butterfly transferred on $SiO_2$/Si substrate. (b) Raman mapping contour based on $E_{2g}$ peak (selected range from 325 $cm^{-1}$ to 375 $cm^{-1}$). (c) PL intensity mapping according to the energy range of 1.95~2.05 eV. (d) Low magnification STEM image of the $WS_2$ butterfly. Red dash line indicates the location of domain boundary between A and B wings. (e) Atomic resolution STEM image of A domain showing 1T phase. (f) STEM image of B domain showing 1H phase.

In order to clarify the optical properties of each $WS_2$ phase, we converted half part of the $WS_2$ butterfly into 1T' phase to realize the coexistence of 1H, 1T, and 1T' phases in a single specimen, therefore, we are able to direct correlate the optical property to each phase accordingly. First of all, the $WS_2$ monolayers grown on sapphire were transferred to a $SiO_2$/Si substrate, and then a rectangular Poly(methyl methacrylate) (PMMA) window was patterned on a $WS_2$ butterfly by using e-beam lithography as shown in the blue contrast area in Fig. 3a. As a result, the $WS_2$ butterfly was divided into four quadrants, the PMMA covered A, B wings and the unmasked C, D wings. Chemical treatment was then performed to the exposed area by using n-butyllithium solution (1.6M in hexane) for 36 hrs in a glove box. After the lithium treatment, the sample was carefully washed with hexane and dried in vacuum, in which the unmasked area treated by lithium ions were all converted into 1T' phase. We characterized the four quadrants of the $WS_2$ butterfly by Raman spectroscopy and PL as shown in Fig. 3d and 3e. The lithium treated area presents clear 1T' Raman features (Fig. 3d green line) and also behaves quenching in the PL spectrum (Fig. 3e green line). However, the A and B domains show again nearly identical Raman feature and PL signals as shown in blue and red lines in the Fig. 3d and 3e, respectively. After the optical characterizations, the corresponding phase of the four quadrants of WS$_2$ butterfly was then confirmed by using atomic resolution STEM imaging. Figure 3b displays the ADF image taken at the A/B domain boundary which verifies the A and B domains as 1T and 1H phases, respectively. On the other hand, the 1T/1T' interface was visualized at the A/C domains shown in Fig. 3c which confirms again the successful 1T to 1T' phase transformation through the lithium chemical treatment at the unmasked area. We further used a state-of-the-art STEM equipped with a monochromator electron source to directly correlate the atomic structure and the exciton absorption of the WS$_2$ phases. [33] Valance loss spectroscopy of the 1H, 1T, and 1T' phase was applied to the same WS$_2$ butterfly specimen (Fig. 3a). More than 60 EELS line scans and 2D maps were acquired crossing the left 1T(A)/1H(B) and 1T(A)/1T'(C) interfaces and also in their individual domains (see Fig. S3 and S4 for more experimental information). Likewise, the exciton adsorption spectra of the 1H and 1T phases WS$_2$ appear in nearly identical semiconducting features as shown the red and blue curves in Fig. 3f, while the 1T' phase (green curve) appears no exciton absorption feature.

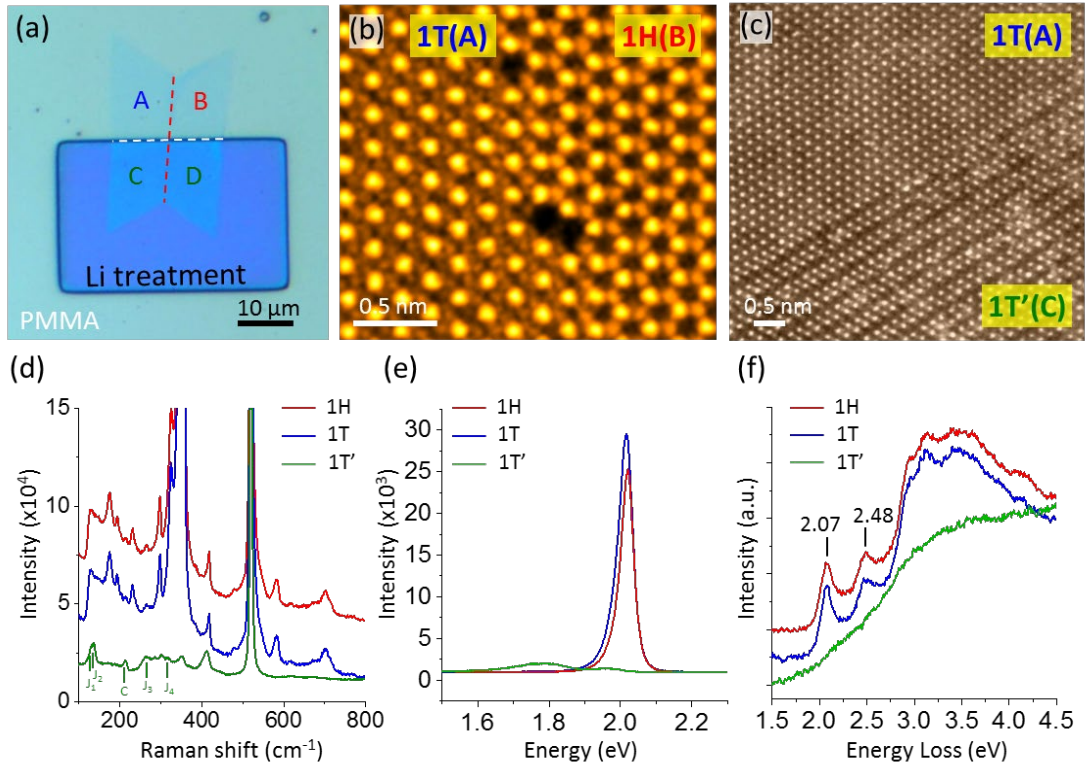

FIG. 3. (a) An optical image of a monolayer $WS_2$ butterfly transferred on a 300nm $SiO_2$/Si substrate. Red dash line indicates the boundary between the left and right wings. The surface was spin coated with PMMA and an e-beam lithographical window was opened in the lower half of the $WS_2$ butterfly. White dash line indicates the unmasked edge on the $WS_2$ butterfly. Four quadrants: A, B, C, and D are assigned to the parts of $WS_2$ butterfly. (b) An ADF image of the $WS_2$ butterfly taken at the A/B domain boundary in (a), where the quadrant-A shows 1T phase, while the quadrant-B shows 1H phase. (c) An ADF image of the $WS_2$ butterfly taken at the A/C interface in (a), where the quadrant-C shows 1T' phase. (d) The Raman spectra, (e) the PL spectra, and (f) the valence loss spectra of $WS_2$ in 1H, 1T, and 1T' phases taken from the $WS_2$ butterfly at room temperature.

Theoretically, the octahedral 1T phase of TMDCs possess partial occupation of the degenerated $4d_{xy/yz/xz}$ orbitals leading to nonzero density of state near Fermi level, as a result, the electronic band structures of 1T-phase $WS_2$ is expected metallic, but the 1H phase is semiconductor [7,9,34]. However, our experimental results indicate that both 1H and 1T $WS_2$ are direct bandgap semiconductor, which is contradicting to the present theory, while the 1T' $WS_2$ shows metal-like behavior with no exciton transition. As charge density wave is an important property in the TMDC families, we have understood that the Peierls distortion and lattice vibration should be considered in the band structure calculation. We have similarly performed DFT calculations to the 1T phase $WS_2$ with strain or possible sulfur lattice distortion or hydrogen absorption, but the results could not fit well to our experimental data. [32] We open this question to reconsider the property of $WS_2$ in 1H, 1T, and 1T' phases and left the calculation for the follow-up work. Correlating the atomic structure and electronic band structure of materials is extremely important to develop the fundamental database of materials properties. It is clear that 1H and 1T phase $WS_2$ present in distinct atomic structures. Although the identical optical and electronic properties of the 1H and 1T phases is beyond one's expectation, this letter is to point out that the current theory may not explain. So we hope that this manuscript may motivate to revisit the perturbation theory and quantum effect on 2D materials with various quasi particles such as CDW and Bardeen–Cooper–Schrieffer superconductors. Although our experimental results are challenging most of the theoretical predictions, this work may hopefully stimulate more research attentions on reconsidering the properties of TMDC polymorphous for applying to further applications.

**Acknowledgement**

YCL and KS acknowledge to the JSPS-KAKENHI (JP16H06333), (18K14119), JSPS A3 Foresight Program, and Kazato Research Encouragement Prize. P.-W.C. appreciates the project support of Taiwan Ministry of Science and Technology: Grants MOST 107-2119-M-007 -011 -MY2 and MOST 106-2628-M-007 -003 -MY3. YCL and KS thank Dr. Silvan Kretschmer and Dr. Arkady V. Krasheninnikov for fruitful discussions and performing preliminary DFT calculations.